\documentclass{article}
\usepackage{spconf,amsmath,graphicx}

\title{A Language Model With Million Context Length For Raw Audio}
\name{Prateek Verma}
\address{
prateekv@stanford.edu\\
Stanford University\\
353, Jane Stanford Way,\\
Stanford, California, 94305
}

\begin{document}
\maketitle
\vspace{2cm}
\begin{abstract}
Modeling long-term dependencies for audio signals is a particularly challenging problem, as even small-time scales yield on the order of a hundred thousand samples. With the recent advent of Transformers, neural architectures became good at modeling dependencies over longer time scales, but they suffered from quadratic constraints to scale them. We propose a generative auto-regressive architecture that can model audio waveforms over quite a large context, greater than 500,000 samples. Our work is adapted to learn time dependencies by learning a latent representation by a CNN front-end, and then learning dependencies over these representations using Transformer encoders, fully trained end-to-end: thereby allowing to learn representations as it deems fit for the next sample. Unlike previous works that compared different time scales to show improvement, we use a standard dataset, with the same number of parameters/context to show improvements. We achieve a state-of-the-art performance as compared to other approaches such as Wavenet, SaSHMI, and Sample-RNN on a standard dataset for modeling long-term structure. This work gives very exciting direction for the field, given improvements in context modeling that can be scaled with more data, as well as potentially better results by using billions/trillions of parameters.

\end{abstract}

\section{Introduction and Related Work}
Modeling long-term dependencies for audio samples is a particularly challenging task, as even smaller time scales yield a hundred thousand samples for typical sampling rates \cite{dieleman2018challenge}. Wavenet architecture first introduced by DeepMind has been a core building block for audio signal processing for a variety of applications. It pioneered modeling audio signals over a long duration, by circumventing various bottlenecks that existed at that time such as vanishing gradients, \cite{hanin2018neural} and the ability of RNNs to model long sequences. Generative modeling for longer sequences is a particularly challenging task, and recently Transformer architectures\cite{vaswani2017attention} have made strong headway in this direction. Transformer architectures have been wildly successful in tasks agnostic of several domains such as genomic sequences \cite{avsec2021effective}, text \cite{brown2020language},  symbolic music \cite{huang2018music}, video\cite{girdhar2019video}, image understanding \cite{dosovitskiy2020image}, audio \cite{verma2021audio,dhariwal2020jukebox}. Even though Transformers have surpassed state of the art in so many diverse fields, they are often constrained by the context size. Wavenet on the other hand has been so widely successful primarily because audio signals are quasi-stationary over smaller time scales which can be easily modeled by convolutional layers. Its performance was further augmented by using conditioning vectors, to guide the generation process, instead of heavy-lifting and learning these dependencies implicitly from the architecture itself. This made wavenet-inspired architectures a core backbone of a variety of audio applications such as speech denoising \cite{rethage2018wavenet}, Text-to-Speech synthesis \cite{shen2018natural}, audio transforms \cite{mor2018universal}. Wavenet has been also used for time-series prediction, as the core architecture, as dilated convolutions give the ability to model trends at micro and macro scales \cite{borovykh2017conditional}. We expect our approach also to hold for similar allied problems such as time-series prediction. The idea of clustering at smaller time scales and then learning dependencies have been explored in non-deep learning approaches \cite{pathak2021cluster} as well as classic deep learning pipelines \cite{schneider2019wav2vec,verma2020framework}. The approach proposed by us also falls in a similar family: We first learn a latent representation, through a CNN front end that can summarize/store important features that are important for a particular signal, and concatenated to a Transformer architectures to learn dependencies across them. We propose this architecture specifically for long term genrative modelling for raw audio. Similar convolutional augmented architectures have been proposed in ASR, and shown improvements in WERs \cite{mohamed2019transformers}. Vector Quantized Variational Autoencoders (VQ-VAEs) \cite{van2017neural} use vector quantization to learn discrete representations of continous signals, that have found success with a variety of applications, such as music synthesis \cite{dhariwal2020jukebox}. However, they are still based on a dilated convolutional based encoder and a decoder to go from the continous signals<-> discrete representations and vice-versa. We do not opt for such architectures, as we do not want any information to get lost, and opt for a classic autoencoder, that can preserve the audio signals in a smaller latent space as much as possible \cite{engel2017neural}. This is where we differ from JukeBox \cite{dhariwal2020jukebox} as well as we use a continuous latent space to learn a latent space for predicting the next sample directly over raw audio instead of predicting a discrete latent code. Neural architectures have been used for learning time-frequency representations from raw audio for problems such as ASR \cite{sainathlearning}, pitch estimation \cite{verma2016frequency}, which shows that they can learn information related to both spectral as well as fundamental frequency associated with any particular chunk of the signal.  However, as shown in \cite{verma2021generative}, Transformer architecture outperforms wavenet in smaller contexts primarily due to the Transformer architecture learning topology as it deems fit as opposed to a fixed topology of dilated convolutions in wavenet. However, it was constrained due to limited context that is plausible with Transformer architecture  The goal of this work is to show we can outperform the performance of wavenet and other work modeling the distribution of raw waveforms, and ingest quite a large context. This work hopefully will serve as a backbone for numerous audio applications where wavenet is involved as a simple drop-in replacement (to do next sample prediction). Further by showing the ability of Transformer architecture to implicitly model distribution over a longer context, this removes the need for conditioning and the model learns various latent representation on its own unlike external conditioning needed in \cite{verma2021generative}.

The contributions of this work are as follows:

i) We produce state-of-the-art results in generative modeling for raw audio. We compare our work for this claim to be, with comparisons against WaveNet, Sample-RNN, and SaSHIMI on the same dataset, with a similar number of parameters, with a fewer number of training steps.

ii) To the best of our knowledge, for raw audio, this is the first work that can do generative modeling for such large contexts. Given that we can model context over 100,000 training examples, this work can further be improved to show improvements over very long contexts even up to a million samples of the past. This is primarily possible due to the inductive bias of audio waveforms to be quasi-stationary over smaller time durations, and the ability of convolutional architectures to compress the contents of the audio signal in a smaller latent space. 
\vspace{-0.6cm}
\begin{figure}[t]
  \centering
  \includegraphics[width=\linewidth]{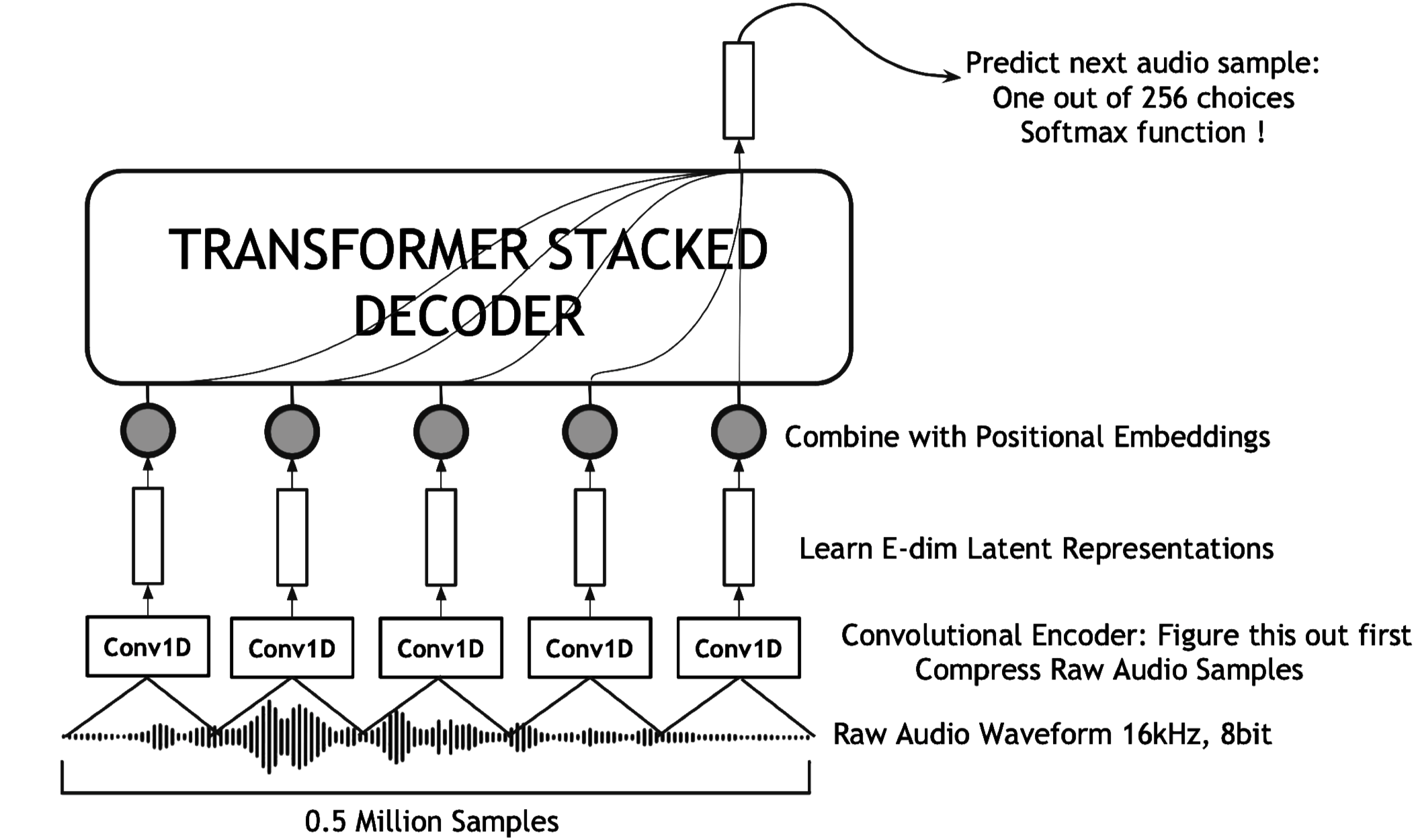}
  \caption{Diagram explaining our architecture. We ingest context of 0.5 million samples to predict the next sample.}
  \label{fig:speech_production}
\end{figure} 
\section{Dataset}
For this work, we have chosen Piano recordings. To evaluate the quantitative results and benchmark our results with other state-of-the-art methods, we use a dataset called YouTube-Mix \cite{goel2022s}. We expect the current results to hold for other popular datasets, such as MAESTRO \cite{hawthorne2018enabling} and speech datasets, primarily since there is no explicit inductive bias present in the model architecture for piano recordings. Secondly, these architecture have shown ubiquity in the domain/datasets to which they are applied to, as seen previous work in WaveNet and Transformer architectures across speech/music/audio signals. However, for the sake of current work, we just focus on reporting the results for piano samples. The choice of the piano is that it has variations present in spectral as well as energy envelops,a wide pitch range, and has a mix of percussive as well as melodic components in addition to being monophonic and polyphonic at the same time. This makes it a good proxy for a large family of audio signals, as well as makes the predictions and modeling at large time scales challenging. We use the same split of training validation and test set as reported in \cite{goel2022s}. The sampling rate chosen was 16kHz, 8-bit resolution and the goal is to predict the next sample (one out of 256 levels for 8-bit signals) given the past. An increased context improves the performance of the model if it can incorporate the context. However, it is often unfairly used to compare different models inaccurately for performance. For a fair comparison of different architectures \cite{goel2022s} with the same setup, \cite{verma2021generative}, for all of the work we use 8s of context giving the architecture learn dependencies across massive time-scales of 128,000 samples for the current work. The choice of 8s is to be consistent in comparing multiple algorithms and methods with our work \cite{goel2022s}. The model trained can be applied to a smaller contexts, and variable lengths as the same convolutional encoder is shared across the time steps, with the same query,key, value matrices learned in Transformer encoder, thereby allowing it to be used with signals other than 8s. As far as data-augmentation is concerned, we do not tinker with the various available data-augmentation techniques to improve performance \cite{schluter2015exploring}. For training, a total of about 10 million unique samples were chosen from the dataset for the next sample prediction during training, which is a fraction of other approaches, making our method robust.

\section{Methodology} In this section we describe the architecture and training details. The model can roughly be divided into three sections, i.e. i) latent representation learning ii) learning dependencies over latent representations iii) prediction of the next sample. We describe each one of the three sub-modules here for the sake of reproducibility and clarity. The goal of generative, auto-regressive architectures is to model the probability distribution of a chunk of waveform.  The joint probability of a waveform, $\textbf{x} = {x_1, x_2, .... x_{T}}$ is modelled as,

\begin{equation}
    p(\textbf{x}) = \prod_{t=0}^{T-1} p(x_{t}|x_0,x_1, x_2, .... x_{t-1})
\end{equation}

Here during training, we minimize the negative log-likelihood of the distribution of the actual one-hot vector and the predicted one. The goal of any architecture that falls in this category is to predict the next sample given the previous context. Of course, we can learn robust latent spaces, by doing a multi-criterion loss as proposed by Deepmind \cite{oord2016wavenet}

\subsection{Learning Latent Representation} As described in the abstract, audio signals are challenging to model at a raw scale. Transformer architectures have been successful in modeling the behavior of audio signals over smaller duration \cite{verma2021generative,child2019generating}. For the case of audio understanding directly from audio waveforms, Transformer architectures were first used by having a front end that can directly be learning a time-frequency representation, and then learn a dependency across these learned representations via attention and feed-forward layers \cite{verma2021audio}. Additionally, for almost all of the audio signals are sparse in the frequency domain, we can easily compress them by a factor of 10x-20x with almost perfect reconstruction. \cite{engel2017neural,haque2018conditional} and \cite{verma2018neural} showed how these latent spaces can be used to encode the information present in the audio signals and can be helpful for the manipulation and synthesis of new timbral spaces. At smaller time scales, audio signals being quasi-stationary larger compression factors are much easier to achieve. For the case of deriving useful information from the signal, as opposed to reconstruction, they have seen success in problems such as speech recognition \cite{schneider2019wav2vec,haque2019audio} and music transcription\cite{hawthorne2018enabling}. The other advantage of such an approach allows us to reduce the size of the attention map, by a factor by which we compress the contents of audio into. A single latent vector of dim-128 per time step is fed to a Transformer. This single latent dim is the compressed version of 2000 audio samples (corresponding to a single time-step of Transformers), thereby giving Transformer the ability to handle quite large sequences with a smaller attention map.  For our case, we divide the audio into chunks of non-overlapping samples of 2000 samples. Each of these chunks (or patches) is passed on a convolutional encoder. It consists of five layers, the first two layers of 256 filters, with the next two being 128 and the final one 32. Strides are alternating between 2 and 3 to subsequently reduce the dimension after every layer, with the kernel size being 7. The final layer tensor is flattened followed by a linear layer to reduce the dimension to 128, to that of the embedding size of Transformers. The weights of the convolutional encoder are shared across all of the chunks.

\subsection{Learning dependencies over Latent Representations} We feed these embeddings to a stacked layer of Transformer modules.  Transformer architecture does not take into account the relative positions of the latent encodings. Sinusoidal positional encodings \cite{vaswani2017attention} are added to each of the latent representations so that the model can explicitly know the position to which each of the latent representations belongs. We use a typical Transformer architecture \cite{verma2021audio}, with a feed-forward dimension of 256, a latent embedding size of 128 and 3 layers. We did not tune the topology as there exist far more combinations given we had access to limited computing resources. Our architecture is minuscule as compared to some of the recent massive architectures used in industry \cite{brown2020language,fedus2021switch}. This leaves room to improve the results of our approach even further. For the context of 128,000 samples, and the chunk size of 2000 samples, we have a total length of an input vector of length 64, and dimension 128 being fed to Transformer architecture. We also randomly drop, with a probability of 0.1 the input sequence and intermediate embedding tokens across time improve robustness.

\subsection{Prediction of the next sample} The transformer blocks yield the same size of the output as the input. The goal of our work is to predict the next 8-bit sample given the past context. Similar to a classification token used in problems such as image classification \cite{dosovitskiy2020image}, we use a single token in the final Transformer layer. This token can be any of the vectors from the final layer of the output and,  can act as a classification token. This is because the attention mechanism can somehow figure out which token is important for classification as inputs are passed on as a list, and the attention mechanism is free to learn importance from any of the 64-length inputs (in our case). However, for the sake of consistency, we choose the 64th element (the last one) to act as a classification token. It has the same size as the embedding of the Transformer, i.e. 128. We now have this as our final latent space that is necessary and sufficient to predict the next sample. As a typical setup, this is passed on through a linear classification head, similar to the manner described in \cite{chen2020simple} to go from a latent space to a classification output. We use a linear layer of dimension 1024 followed by a 256 dim layer to produce the output space same as the number of possible states (256 states for 8-bit audio signals). The dropout rate chosen for the linear classification head is 0.2, with softmax activation get probability distribution of next sample. 

\subsection{Training Details} The models were trained with a batch size of 40 on four A100 GPUs using Adam \cite{kingma2014adam} with the loss function as minimizing the cross-entropy between the actual distribution of the next sample and that predicted by wavenet. We use the initial learning rate of 1e-4 for about 250,000 training steps, followed by 0.5*1e-5 for the remaining till we saw the held-out validation loss plateauing, of about 2 epochs in a total of 10 million samples per epoch. We expect the results to improvewith more training, and incorporation of data-augmentation strategies such as changing the volume dynamics, pitch shifts. Dropout rate of 0.2 was chosen to improve generalization on feed-forward and convolutional layers. 

\section{Results and Discussion} For reporting the results, we compare our work against the recent state-of-the-art neural architectures i.e. Wavenet \cite{oord2016wavenet}, Sample-RNN \cite{mehri2016samplernn} and recently introduced SaSHMI \cite{goel2022s}. We primarily evaluate these approaches on negative-log likelihood values (NLL=). Similar to the reasons described in \cite{verma2021generative}, the choice is because we want to disentangle the application-specific results from how well the model does in the next step prediction task. We expect that a model that does well in predicting the next sample, would generalize well for other tasks including unconditioned generation. The wavenet baseline used was from \cite{goel2022s}. Also, \cite{verma2021generative}, compared smaller time scales with similar context and exact same number of parameters of a wavenet with that of a Transformer baseline to show the superiority of attention based architectures. Table I shows our results against other state-of-the-art architectures on the YouTubeMix piano dataset.
\vspace{-0.3cm}
\begin{table}[ht]
  \caption{Comparison of the negative-log likelihood (NLL) scores for our architecture with state-of-the-art methods. We achieve the same with similar parameters and most importantly beat them with similar context. A lower score is better. We use context the for WaveNet, and SampleRNN namely 1024 and 4092 respectively. For Comparision with SaShimi, we use a context of 128,000 samples to be consistent}
	\centering
	\begin{tabular}{|c|c|c|}
		\hline
		Neural Model Architecture & \#of param &NLL score\\\hline
		Wavenet \cite{goel2022s,oord2016wavenet} & 4.24M  & 1.45\\
		SampleRNN: 3 tier \cite{goel2022s,mehri2016samplernn} & 35.03M  & 1.72\\
		SaShimi: 6 Layers  \cite{goel2022s} & 3.13M  & 1.32 \\\hline
		Ours: 3 Layers 128k context & 3.14M & \textbf{1.26} \\
		Ours: 3 Layers 0.5 M context & 3.14M & \textbf{1.19} \\\hline
		
	\end{tabular}
	\label{tab:example}
\end{table}

As previously noted, these results were trained on about 10 million samples in our case. We can expect these results to improve further if we can ingest more training data and further pre-train on large-scale audio recordings. The goal of this work was to have a fair comparison, and for that sake, and due to the availability of computing resources, we could not go beyond what is reported. To give a context, we had distinct training samples of about 10 million points, sampled uniformly over roughly 4 hours of training split. This was out of the possible ~ 57M possible samples available in 4 hours of training split. We expect these results to improve further by having a bigger diverse, larger dataset as reported in works such as \cite{ellis1999size,radford2018improving}. 

\section{Conclusions and Future Work}

We have successfully shown how to do auto-regressive modeling for raw audio with really long context inputs. Further, to the best of our knowledge, there does not exist any past work that can incorporate contexts that are so long-up to a million samples and show that they improve the performance. We believe the approach of learning a latent representation that can incorporate frequency as well as timing information, similar to learning a Fourier representation on smaller time scales and using attention-based Transformer architecture has applications in other domains too. We are also excited to have shown significant improvements with a minuscule number of parameters. Further competing on modeling long context (greater than 500k samples) is an interesting research direction as opposed to chasing model size and parameters and we expect this work to be a stepping stone in that direction. We also envision the current architecture to have significantly improve performance of several time-series prediction problems.

\bibliographystyle{IEEEbib}
\bibliography{refs}

\end{document}